\documentclass{iaus}
\usepackage{graphicx}

\title[The macroturbulence\,--\,pulsation conection] 
{Is macroturbulence in OB Sgs related to pulsations?}

\author[S. Sim\'on-D\'iaz et al.]   
{S. Sim\'on-D\'iaz$^{1,2}$,
A. Herrero$^{1,2}$,
K. Uytterhoeven$^{3}$
N. Castro$^{1,2}$
C. Aerts$^{4,5}$
\and J. Puls$^{6}$
}

\affiliation{$^1$Instituto de Astrof\'\i sica de Canarias, E-38200 
           La Laguna, Tenerife, Spain.
$^2$Departamento de Astrof\'isica, Universidad de La Laguna, E-38205 
           La Laguna, Tenerife, Spain.
$^3$ Laboratoire AIM, CEA/DSM-CNRS-Universit\'e Paris Diderot; 
CEA, IRFU, SAp, centre de Saclay, 91191, Gif-sur-Yvette, France.
$^4$ Instituut voor Sterrenkunde, Katholieke Universiteit Leuven, Celestijnenlaan 200D, 3001 Leuven, Belgium. 
$^5$ IMAPP, Department of Astrophysics, Radboud University Nijmegen, PO Box 9010, 6500 GL Nijmegen, the Netherlands.
$^6$ Universit\"atssternwarte M\"unchen, Scheinerstr. 1, 81679 M\"unchen, Germany.
\\
email: {\tt ssimon@iac.es}}

\pubyear{2010}
\volume{272}  
\pagerange{119--126}
\jname{Active OB stars: structure, evolution, mass loss and critical limits}
\editors{C. Neiner, G. Wade, G. Meynet \& G. Peters, eds.}
\begin{document}

\maketitle

\begin{abstract}
As part of a long term observational project, we are investigating the macroturbulent
broadening in O and B supergiants (Sgs) and its possible connection with
spectroscopic variability phenomena and stellar oscillations. We 
present the first results of our project, namely
firm observational evidence for a strong correlation between the
extra broadening and photospheric line-profile variations in a
sample of 13 Sgs with spectral types ranging from O9.5 to B8.

\end{abstract}

\firstsection 
\section{Introduction}

The presence of an important extra line-broadening mechanism (in addition to the rotational 
broadening and usually called macroturbulence) affecting the spectra of O and B Sgs is 
well established observationally (see Sim\'on-D\'iaz et al. 2010, and references therein).
\cite[Lucy (1976)]{Luc76} postulated that this extra\,broadening may be identified 
with surface motions generated by the superposition of numerous non-radial oscillations.  
More recently, \cite[Aerts et al. (2009)]{Aer09} computed time\,series of line profiles for 
evolved massive stars broadened by rotation and hundreds of low amplitude non-radial gravity 
mode oscillations and showed that the resulting profiles could mimic the observed ones. 
Stellar oscillations are a plausible explanation for the extra\,broadening in O and B Sgs, but
this hyphotesis needs to be observationally confirmed.

\section{The macroturbulence\,--\,LPV connection}

As a first step, in \cite[Sim\'on-D\'iaz et al. (2010)]{Sim10}, we investigated the possible connection 
between the macroturbulent broadening and the presence and temporal behaviour of 
line-profile variations (LPVs) in a sample of 11 late-O and early-B Sgs, 2 late B-Sgs, 
and 2 late-O, early-B dwarfs. To this aim, we obtained and analyzed time\,series of 
high resolution (R\,$\sim$\,46000), high S/N spectra obtained with FIES@NOT in two
observing runs. We applied the Fourier transform (\cite[Gray 1976]{Gra76}) and the 
goodness-of-fit techniques to disentangle and measure the contributions from 
rotational ($v$\,sin$i$) and macroturbulent ($\Theta_{\rm RT}$) broadening 
to the Si\,{\sc iii}\,4567 and/or the O\,{\sc iii}\,5592 line profiles. 
We quantified the LPVs in these lines by means of the first, 
$\langle v \rangle$, and third, $\langle v^3 \rangle$, normalized velocity 
moments of the line. These moments 
are related to the centroid velocity and the skewness of the line profile, 
respectively, and are well suited to investigate whether an observed line profile is subject to 
time-dependent line asymmetry, as expected in the case of a pulsating star.

\begin{figure}[t!]
\begin{minipage}[l]{0.50\textwidth}
\begin{center}
\includegraphics[width=7.5cm,angle=90]{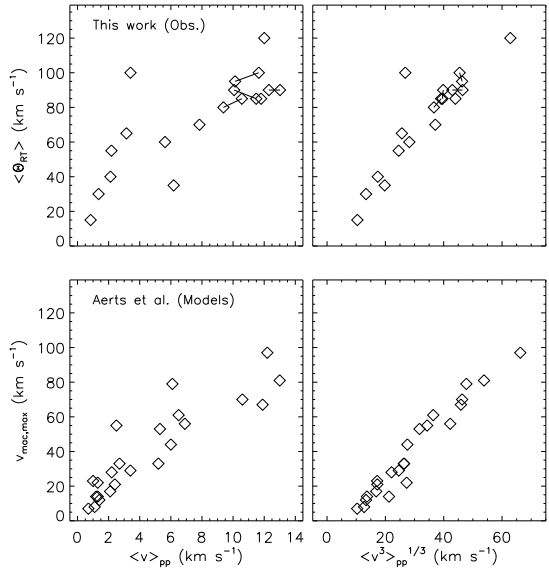}
\end{center}
\end{minipage}  
\begin{minipage}[l]{0.07\textwidth}
\
\end{minipage}  
\begin{minipage}[r]{0.40\textwidth}
\begin{center}
      \caption{(Top) Empirical relations between the average size of the macroturbulent broadening
	  ($\langle\Theta_{\rm RT}\rangle$) and the peak-to-peak amplitude of the first and third moments of the
	  line profile. Solid lines connect results from four stars observed in both campaigns.
(Bottom) similar plots with data
  from Table 1 in \cite[Aerts et al. (2009)]{Aer09}, based on simulations of line profiles broadened
  by rotation and by hundreds of low amplitude non-radial gravity mode
  pulsations. The simulations lead to clear trends which are compatible with spectroscopic
observations.}
         \label{f2}
\end{center}
\end{minipage} 
\end{figure}

We found a clear positive correlation between the average size of the macroturbulent 
broadening, $\langle \Theta_{\rm RT} \rangle$, and the peak-to-peak amplitude of 
$\langle v \rangle$ and $\langle v^3 \rangle$ variations (see Fig. \ref{f2}). To our 
knowledge, this is the {\em first clear observational evidence for a connection between 
extra broadening and LPVs in early B and late O Sgs.}

\section{Is macroturbulent broadening in OB-Sgs caused by pulsations?}

Non-radial oscillations have been often suggested as the origin of LPVs and
photosperic lines in OB Sgs; however, a firm confirmation (by means of a 
rigorous seismic analysis) has not been achieved yet. From a theoretical 
point of view, \cite[Saio et al. (2006)]{Sai06} showed that g-modes can 
be excited in massive post-main sequence stars, as the g-modes are 
reflected at the convective zone associated with the H-burning shell. 
\cite{Lef07} presented observational evidence of g-mode instabilities 
in a sample of photometrically variable B\,Sgs from the location of the 
stars in the (log\,T$_{\rm eff}$, log\,$g$)-diagram.

These results, along with our observational confirmation of a tight
connection between macroturbulent broadening and parameters describing observed 
LPVs render stellar oscillations the
most probable physical origin of macroturbulent broadening in B\,Sgs; however,
it is too premature to consider them as the only physical phenomenon to explain
the unknown broadening.

\end{document}